\documentclass[aps,prx,reprint,preprintnumbers,superscriptaddress,nofootinbib,longbibliography,floatfix]{revtex4-2}
\pdfoutput=1

\usepackage{rotating}
\usepackage{array}
\usepackage{amsmath}
\usepackage[normalem]{ulem}
\usepackage{slashed}
\usepackage{booktabs}
\usepackage[pdftex,table]{xcolor}
\usepackage{units}
\usepackage{xfrac}
\usepackage{mathtools}
\usepackage{empheq}
\usepackage[]{units}
\usepackage{multirow}
\usepackage{amssymb}
\usepackage{url}
\usepackage{comment}
\usepackage{physics}
\usepackage{color,soul}
\usepackage{bbm}
\usepackage[caption=false]{subfig}

\usepackage{hyperref}
\hypersetup{
  colorlinks=true,
  citecolor=blue,
  linkcolor=blue,
  urlcolor=blue
}

\newcommand{\cgan}{\textsc{CaloGAN}}
\newcommand{\cf}{\textsc{CaloFlow}}
\newcommand\geant{\textsc{Geant}4}
\begin{document}
\preprint{HEPHY-ML-24-01}
\title{Unifying Simulation and Inference with Normalizing Flows}

\author{Haoxing Du}
\email{haoxingdu@gmail.com}
\affiliation{Department of Physics, University of California, Berkeley, CA 94720, USA}

\author{Claudius Krause}
\email{Claudius.Krause@oeaw.ac.at}
\affiliation{Institut f\"ur Theoretische Physik, Universit\"at Heidelberg, Philosophenweg 12, 69120 Heidelberg, Germany}
\affiliation{Institute of High Energy Physics (HEPHY), Austrian Academy of Sciences (OeAW), Dominikanerbastei 16, A-1010 Vienna, Austria}

\author{Vinicius Mikuni}
\email{vmikuni@lbl.gov}
\affiliation{National Energy Research Scientific Computing Center, Berkeley Lab, Berkeley, CA 94720, USA}

\author{Benjamin Nachman}
\email{bpnachman@lbl.gov}
\affiliation{Physics Division, Lawrence Berkeley National Laboratory, Berkeley, CA 94720, USA}
\affiliation{Berkeley Institute for Data Science, University of California, Berkeley, CA 94720, USA}

\author{Ian Pang}
\email{ian.pang@physics.rutgers.edu}
\affiliation{NHETC, Department of Physics and Astronomy, Rutgers University, Piscataway, NJ 08854, USA}

\author{David Shih}
\email{shih@physics.rutgers.edu}
\affiliation{NHETC, Department of Physics and Astronomy, Rutgers University, Piscataway, NJ 08854, USA}

\begin{abstract}
There have been many applications of deep neural networks to detector calibrations and a growing number of studies that propose deep generative models as automated fast detector simulators. We show that these two tasks can be unified by using maximum likelihood estimation (MLE) from conditional generative models for energy regression. Unlike direct regression techniques, the MLE approach is prior-independent and non-Gaussian resolutions can be determined from the shape of the likelihood near the maximum. Using an ATLAS-like calorimeter simulation, we demonstrate this concept in the context of calorimeter energy calibration.
\end{abstract}

\maketitle

\section{Introduction}

Detector calibrations are one of the most important and foundational tasks in experimental physics.  In particle and nuclear physics, the largest calibration step is usually a simulation-based\footnote{The residual correction to account for differences between data and simulation is usually smaller.  Methods in this paper may also be applicable to data-based corrections in certain cases.} correction to ensure that the reported properties of the measured particles are unbiased.  Detector simulations are complex and accurate, but can only run forward in time and the resolutions are non-trivial.  Combined, these properties mean that we cannot simply invert the simulation to predict true quantities given measured ones.  

This is particularly acute for highly-segmented detectors where many individual channels are activated for a single particle as the measured phase space can be high dimensional and complex.  A common example of this setting is calorimeter reconstruction. Unlike tracking detectors that aim to minimally disrupt the trajectory of a particle, calorimeters are designed to stop particles and the resulting showers inside dense materials are challenging to parse and are highly stochastic.  Traditional calibration methods are optimized using relatively low-dimensional summary statistics.  Such calibrations have enabled many science results, but the ultimate precision cannot be reached until we utilize all of the available low-level information for event reconstruction.

Machine learning provides a set of tools that can analyze hadronic final states holistically to achieve the best precision. Recent examples supporting this claim include the energy calibrations of single hadrons~\cite{deOliveira:2018hva,CMS:2017jpq,ATL-PHYS-PUB-2020-018,Neubuser:2021uui,Akchurin:2021afn,Kieseler:2021jxc,Akchurin:2021ahx,ATLAS:2022geo,Qasim:2022rww,Acosta:2023nuw,ATLAS:2023zzb}, jets~\cite{Haake:2018hqn,CMS:2019uxx,Cheong:2019upg,Haake:2019pqd,CMS-DP-2021-017,Gambhir:2022gua,Gambhir:2022dut,ATLAS:2023tyv,ATLAS:2023zca,ALICE:2023waz},  and global event properties~\cite{Diefenthaler:2021rdj,Arratia:2021tsq,Leigh:2022lpn,Raine:2023fko} at colliders.
These techniques can automatically make use of finer segmentation to improve the resolution of reconstructed particle energies~\cite{Neubuser:2021uui,Acosta:2023nuw}.  While machine learning has broader utility than reconstructing individual hadron showers within a calorimeter, we focus on this case because it is particularly challenging and a critical component of event reconstruction at the Large Hadron Collider and elsewhere. The process of inverting detector simulation is closely related to
unfolding. Various unfolding methods have been developed and are actively utilized in experimental research. For comprehensive reviews, refer to Refs.~\cite{Cowan:2002in,Blobel:2011fih,doi:https://doi.org/10.1002/9783527653416.ch6,Brenner:2019lmf}, and for the most commonly used unfolding algorithms, see Refs.~\cite{DAgostini:1994fjx,Hocker:1995kb,Schmitt:2012kp}. The main difference between unfolding and calibration is that unfolding aims to correct distributions, while calibration aims to correct individual objects/events. Recently, there have been several proposed ML-based unfolding methods that are able to achieve remarkable unfolding performance. See Refs.~\cite{Datta:2018mwd,Andreassen:2019cjw,Bellagente:2019uyp,Bellagente:2020piv,Vandegar:2020yvw,Howard:2021pos,Andreassen:2021zzk,Leigh:2022lpn,Backes:2022vmn,Shmakov:2023kjj,Ackerschott:2023nax,Diefenbacher:2023wec,Butter:2023ira,Shmakov:2024gkd,Huetsch:2024quz} for examples of ML-based unfolding methods.

While existing machine learning methods are promising, most approaches have at least two undesirable features: they are only point estimates and are prior-dependent. Firstly, most approaches produce a single estimate without quantifying `uncertainties'. In particle/
nuclear physics, the spread of the difference between the inferred estimates and the true value is usually called the \textit{resolution}, even though it is part of Uncertainty Quantification (UQ) in the broader machine learning literature.
 Secondly, techniques that employ standard machine learning regression tools are not \textit{universal}: their performance depends on the distribution of examples (the `prior', e.g. uniform in energy) used during training. In fact, this prior dependence often results in a large calibration bias.

Over the last years, some solutions to these two central challenges have been proposed. Generalized numerical inversion~\cite{Cukierman:2016dkb,ATL-PHYS-PUB-2020-001,ATL-PHYS-PUB-2018-013} is a prior-independent calibration approach, although it does not scale well to many output dimensions. The Gaussian Ansatz~\cite{Gambhir:2022gua,Gambhir:2022dut} is a maximum likelihood estimator, and thus prior-independent (more in Sec.~\ref{sec:methods}). For UQ, loss functions can be modified to estimate quantiles in addition to just the mean, median or mode~\cite{CMS:2019uxx,Cheong:2019upg} and deep generative models can estimate the exact resolution function~\cite{Leigh:2022lpn,Raine:2023fko}. In the CMS Run 1 Higgs analysis~\cite{CMS:2013btf}, the uncertainty of the photon energy regressed by a boosted decision tree (BDT) is estimated by training an additional BDT. The Gaussian Ansatz produces an estimate of the resolution and so is currently the only method that is both prior-independent and that goes beyond a point estimate.  We note that it is also possible to use Bayesian versions~\cite{mackay1995probable,neal2012bayesian,gal2016uncertainty,Kasieczka:2020vlh,Bellagente:2021yyh} of neural networks to go beyond point estimates by providing the uncertainty of the output. However, the uncertainty would still be prior-dependent. 

In this paper, we propose a prior-independent method for detector calibrations that produces per-shower resolution estimates. The approach is based on deep generative models with access to the explicit likelihood.  In particular, we use normalizing flows\footnote{It is also possible to extract the likelihood in many dimensions from a diffusion model~\cite{Mikuni:2023tok} or if the inputs are discretized, from an autoregressive model~\cite{Finke:2023veq}.}~\cite{tabak_flows,dinh2014nice,razende_flows,dinh2016density,flows_review,CNFs}. These machine learning models are invertible functions with a tractable Jacobian so that one can be used for both sampling and density estimation. The core idea behind a normalizing flow is that one starts with a simple random variable with a known probability density (e.g. a Gaussian) and then applies a series of transformations with the tractable Jacobian. This results in a complex probability density that can also be computed via the change of variables formula. The probability density can be made conditional by letting each transformation depend on the conditional quantity. Our approach serves as an alternative to other methods based on maximum likelihood estimation.
%
%\nocite{Paganini:2017hrr,Paganini:2017dwg,deOliveira:2017rwa,Erdmann:2018kuh,Erdmann:2018jxd,ATL-SOFT-PUB-2018-001,Belayneh:2019vyx,Vallecorsa:2019ked,SHiP:2019gcl,Chekalina:2018hxi,Carminati:2018khv,Vallecorsa:2018zco,Musella:2018rdi,Deja:2019vcv,ATLAS:2022jhk,ATLAS:2021pzo,ATLAS:2022jhk,Buhmann:2021lxj,Buhmann:2021caf,Krause:2021ilc,Krause:2021wez,Mikuni:2022xry,Buckley:2023rez,Krause:2022jna,Diefenbacher:2023vsw,Cresswell:2022tof, Liu:2023lnn,Diefenbacher:2023prl,Buhmann:2023bwk,Mikuni:2023tqg,Acosta:2023zik}
%

Normalizing flows are state-of-the-art as calorimeter simulation surrogate models~\cite{Krause:2021ilc,Krause:2021wez,Krause:2022jna,Buckley:2023rez, Diefenbacher:2023vsw,Xu:2023xdc,Pang:2023wfx,Ernst:2023qvn}. Additionally, normalizing flows have shown similar good performance in other tasks in high-energy physics~\cite{Nachman:2020lpy,Gao:2020vdv,Bothmann:2020ywa,Gao:2020zvv,Bellagente:2020piv,Stienen:2020gns,Bieringer:2020tnw,Bellagente:2021yyh, Hallin:2021wme,Bister:2021arb,Butter:2021csz, Winterhalder:2021ngy, Butter:2022lkf,Verheyen:2022tov, Leigh:2022lpn, Butter:2022vkj,Hallin:2022eoq,Heimel:2022wyj,Backes:2022vmn,Leigh:2022lpn,Raine:2023fko,Sengupta:2023xqy, Ackerschott:2023nax,Heimel:2023mvw,Heimel:2023ngj,Bierlich:2023zzd,Das:2023bcj}. In this paper, we unify simulation and inference by showing how models trained for generation can be reused for calibration. As a demonstration of this approach, we use \cf\ \cite{Krause:2021ilc,Krause:2021wez}, a normalizing flow-based calorimeter surrogate model. This machine learning model is trained on single-pion showers from an extended version~\cite{du_2024_11073232} of the \cgan~dataset~\cite{nachman2017electromagnetic,Paganini:2017hrr,Paganini:2017dwg} that now includes both a sampling calorimeter setup~\cite{Krause:2023uww} and a hadronic component.  The calibration task is to predict the incident pion energy given the distribution of energies recorded in the cells of the calorimeter. Importantly, we demonstrate the following key advantages of our approach:
\begin{enumerate}

    \item \textbf{Zero-shot calibration:} \\ 
    Normalizing flow-based models trained for surrogate modelling can
be reused for calibration by repurposing the probability
density as a likelihood without any additional model retrainings.

    \item \textbf{Access to per-shower resolution:}\\
    With access to the complete likelihood, we can compute the exact resolution function. This allows us to obtain per-shower resolution estimates which give us additional information about how close the predicted energy is to the true energy. Furthermore, it enables us to estimate asymmetries in the resolution function.
    \item \textbf{Less biased calibration:}\\
    We achieve a smaller calibration bias compared to a typical direct regression approach for the calorimeter setup considered in this work.

\end{enumerate}
  
This paper is organized as follows. Machine learning-based calibration methods are introduced in Sec.~\ref{sec:methods}. Here we also provide precise definitions for \textit{bias} and \textit{resolution} used in this paper. In Sec.~\ref{sec:caloexample}, we summarize the \cf~algorithm and discuss the results of the calorimeter example which demonstrate the key advantages listed above. The paper ends with conclusions and outlook in Sec.~\ref{sec:conclusions}. We collect some details on the used network architectures in an appendix.

\section{Methods}
\label{sec:methods}

Given samples from a forward model (usually a physics-based simulation) $X\sim p_{X|Z}(x|z)$ for measured values $X$ and true values $Z$, the goal of calibration is to estimate $Z$ from $X$.  In our notation, capital letters represent random variables and lower case letters represent realizations of those random variables.  Both the measured and true values can be multidimensional, although we will focus on the common case where $Z$ is one dimensional and $X\in\mathbb{R}^N$.

The usual assumption when deriving a simulation-based calibration is that the detector response is correct and universal, i.e. $p_{X|Z}^\text{train}(x|z)=p_{X|Z}^\text{test}(x|z)$ for different physics processes (i.e.~different $p(z)$) or for actual data. 

There are two important quantities related to calibration performance --- the bias and the resolution. 

\begin{enumerate}
    \item The \textbf{bias} of a calibration is the deviation between
the central tendency of the inferred estimate $\hat z$ and the
true reference value $z$. Any measure of central tendency
can be used to measure closure, such as the median or
mode. In this paper, we will focus on the mode\footnote{Usually bias refers to the mean, but experimentalists typically calibrate to the mode.} of the
inferred estimates given a fixed $z$ value. To compute the mode in practice, for a given $z$, we look at $K$ measured values $x_1, x_2 \dots x_K$ corresponding to that $z$. Each measured value $x_i$ undergoes calibration to derive an inferred estimate $\hat z_i$. The mode is subsequently determined based on the collection of inferred estimates $\{\hat z_1,\hat z_2,\dots, \hat z_K\}$ (see steps in Sec.~\ref{sec:bias}).
\item The \textbf{resolution} is the spread of the difference between the inferred estimates and the true value. In this work, we have two relevant definitions:

\begin{itemize}
\item \textbf{Full resolution:} For fixed $z$, it is defined as half of the 68\% confidence interval about the mode determined based on a collection of inferred estimates.

\item \textbf{Per-event resolution:} For each $x$, it is defined as half of the 68\% confidence interval of the corresponding maximum likelihood estimate (see Sec.~\ref{sec:MLE}). 
\end{itemize}
The full resolution is not to be confused with the per-event resolution. The full resolution is defined for a collection of inferred estimates, whereas the per-event resolution is defined for each inferred estimate.
\end{enumerate} 

\subsection{Direct Regression}

Most proposals for deep learning-based calibration directly regress $Z$ from $X$ using a loss function like the mean-squared error (MSE):

\begin{align}
\label{eq:mse}
    L[f]=\sum_i (f_{\rm MSE}(x_i)-z_i)^2\,,
\end{align}
for a neural network $f_{\rm MSE}$.  Using the calculus of variations, one can show that with enough training data and sufficiently flexible neural network architecture and training protocol, the solution to Eq.~\ref{eq:mse} is the average value of $Z$ given $X=x$:

\begin{align}
f_{\rm MSE}(x) &= \langle Z|X=x\rangle \\
    &=\int dz\,z\,p_{Z|X}^\text{train}(z|x)\\\label{eq:mse2}
    &=\int dz\, z\, p_{X|Z}^\text{train}(x|z)\,\frac{p_Z^\text{train}(z)}{p_X^\text{train}(x)}\,.
\end{align}
Note that $f_{\rm MSE}$ only gives a point estimate of $z$ for each $x$ and does not have access to per-event resolutions. In other words, there is no calibration uncertainty quantified by $f_{\rm MSE}$.

For a given $z$, the bias can be computed as the deviation between $z$ and the mode of a collection of $f_{\rm MSE}(x_i)$, and the full resolution at $z$ is then computed based on the 68\% confidence interval about the mode. Note that the bias is defined as a function of $z$.\footnote{Alternatively, defining the bias as a function of the inferred values would result in the bias being dependent on the prior used during testing (i.e.~$p^{\rm test}_Z(z)$). This bias also cannot be corrected since we do not know a priori what $p^{\rm test}_Z(z)$ is used during calibration.} Hence, we cannot correct for this bias as we do not have access to $z$ while performing the calibration. 

The challenge with Eq.~\ref{eq:mse2} is that it depends on $p_Z^\text{train}$ even when we assume that the detector response is universal. Unsurprisingly then, the bias of direct regression also depends on the training dataset. In Ref.~\cite{Gambhir:2022dut}, the prior dependence of MSE-based calibration is shown to be the main source of large calibration bias.

\subsection{Maximum Likelihood Inference}
\label{sec:MLE}
Maximum likelihood estimation (MLE) involves finding the $Z$ that maximizes the likelihood of the data,
\begin{equation}
\label{eq:mle}
  \hat{Z} = \underset{z}{\text{argmax}}\, p_{X|Z}(X | z)\,.
\end{equation}
Since we assume that $p_{X|Z}$ is universal and Eq.~\ref{eq:mle} only depends on this likelihood, then $\hat{Z}$ is universal. We can go beyond the point estimate and use the likelihood function $p_{X|Z}$ to obtain the per-event resolution for each maximum likelihood estimate $\hat z$. The access to per-event resolutions is a major advantage of MLE over direct regression in calibration. Examples showcasing this advantage are discussed in Sec.~\ref{sec:resolution}.

Similar to the direct regression case, the bias and full resolution at a fixed $z$ value can be defined based on the mode and 68\% confidence interval about the mode.

Although the MLE-based calibration is always prior-independent, the prior independence does not guarantee that the method is always unbiased. Only in certain scenarios (e.g.~1D Gaussian noise model such that $p_{X|Z}(x|z) \sim N(z,\sigma)$) can the MLE calibration be shown to be unbiased. Certain detector responses $p_{X|Z}$ may result in biased calibration. In general, prior independence is a necessary but insufficient condition for unbiased calibration. Even so, we show in Sec.~\ref{sec:bias} that MLE-based calibration results in a smaller bias compared to direct regression for an ATLAS-like calorimeter setup. This suggests that in our application, the dependence on training prior is a bigger contributing factor to the bias than the detector response. 

Also, the challenge with MLE is that we usually do not know $p_{X|Z}$ explicitly.  Nevertheless, we are able to sample from this conditional density by running a simulation. In this work, we instead aim to learn the entire likelihood using a neural network --- an approach sometimes referred to as {\it neural likelihood estimation} in the simulation-based inference literature~\cite{papamakarios2019sequential,papamakarios2019neural,cranmer2020frontier,dirmeier2023simulation}.  This is substantially more work, but we observe that the work may already be done: given a fast simulation based on neural networks with access to the likelihood, we can use it for calibration in addition to generation without requiring any additional retraining.  

Our tool of choice is the normalizing flow (NF). NFs are neural networks that are optimized using maximum likelihood estimation.  With access to an estimate of the full likelihood, we can study both Gaussian and non-Gaussian aspects of the resolution.

\section{Regressing particle incident energy}
\label{sec:caloexample}

In this section, we compare the performance of direct regression and NF in regressing the particle incident energy from the resulting calorimeter shower information.

\subsection{Dataset}
\label{sec:dataset}

The events are $\pi^+$ calorimeter showers from a new version~\cite{du_2024_11073232} of the \cgan~dataset which we generated with \geant~\cite{Agostinelli:2002hh,1610988,ALLISON2016186} for this study. We generated 100k showers with $\pi^+$ incident energy uniformly distributed in the range [1,100] GeV. Additionally, we generated 100k showers with incident energy log-uniformly distributed in the same range. The calorimeter setup used in the original dataset only included the electromagnetic calorimeter (ECAL). In this new dataset, we also include the hadronic calorimeter (HCAL) which is positioned behind the ECAL. Also, the original dataset included energy contributions from both active and inactive calorimeter layers, whereas this new \textit{sampling} calorimeter dataset only includes energy contributions from the active layers as would be available in practice.

The ECAL is a three-layer sampling calorimeter cube with 480~mm side-length that is inspired by the ATLAS liquid argon (LAr) electromagnetic calorimeter~\cite{CERN-LHCC-96-041}. For the ECAL, the active
material is LAr and the absorber material is lead (Pb). Note that the ECAL used here is identical to the one used in Ref.~\cite{Krause:2023uww}. However, photon (instead of $\pi^+$) showers were considered in that study. The HCAL is a three-layer sampling calorimeter cube with 2000~mm side-length located behind the ECAL. For the HCAL, the active
material is LAr and the absorber material is tungsten (W). The sampling fractions of the ECAL and HCAL are $\sim$ 20\% and $\sim$ 1.3\%, respectively. 
The calorimeter showers are represented as three-dimensional images that are binned in position space. 

\begin{figure*}[!ht]
\includegraphics[width=0.66\columnwidth]{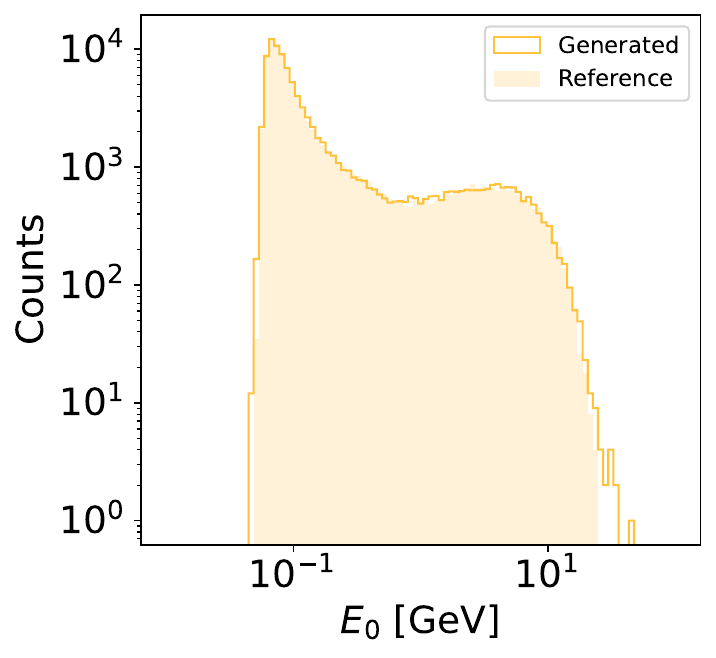}\includegraphics[width=0.66\columnwidth]{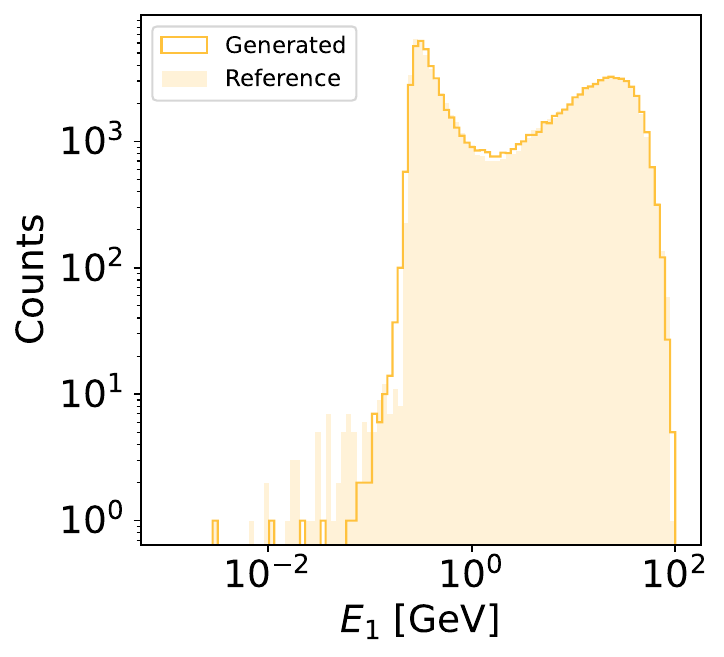}\includegraphics[width=0.66\columnwidth]{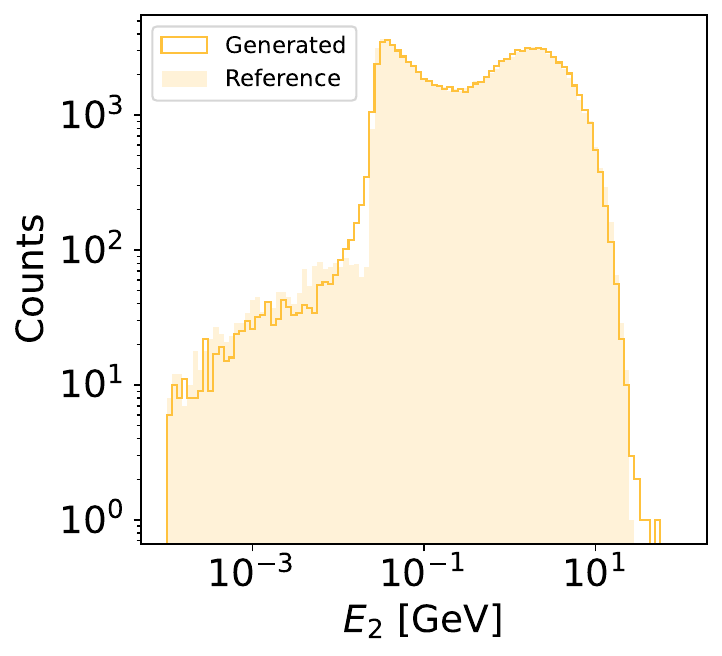}\\
\includegraphics[width=0.66\columnwidth]{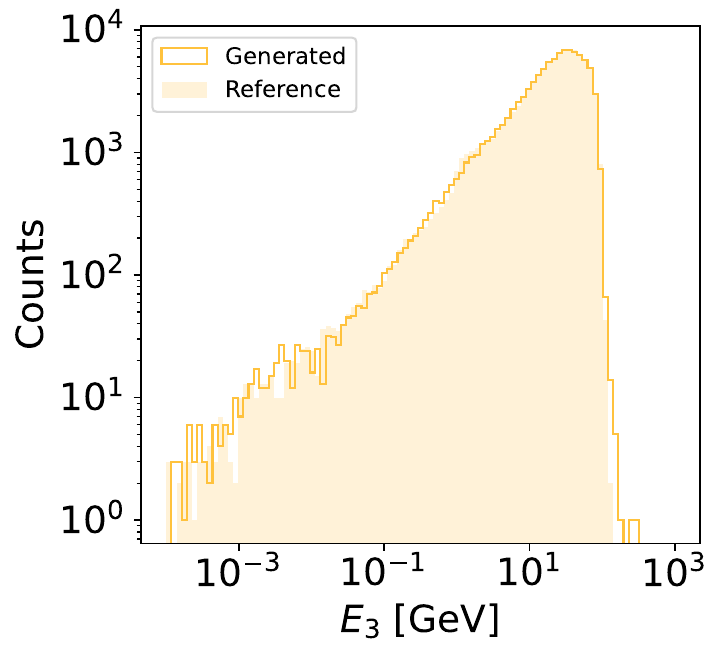} 
\includegraphics[width=0.66\columnwidth]{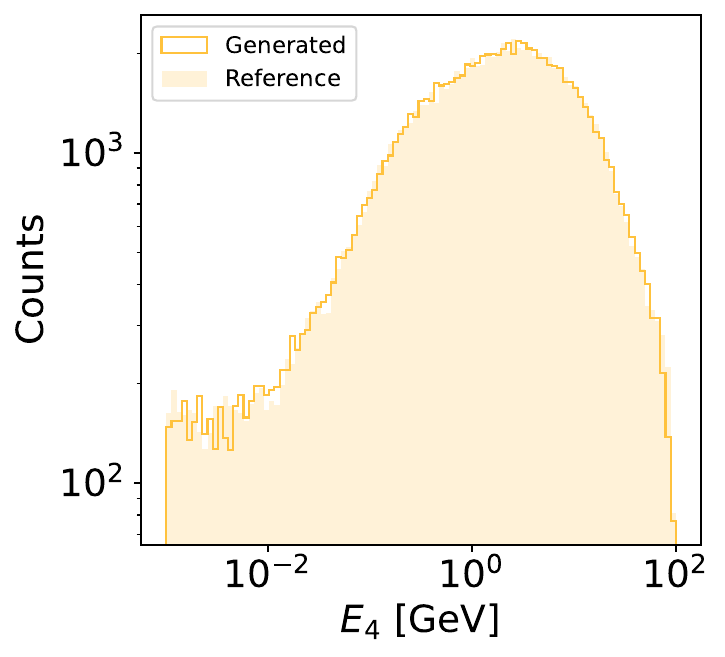}\includegraphics[width=0.66\columnwidth]{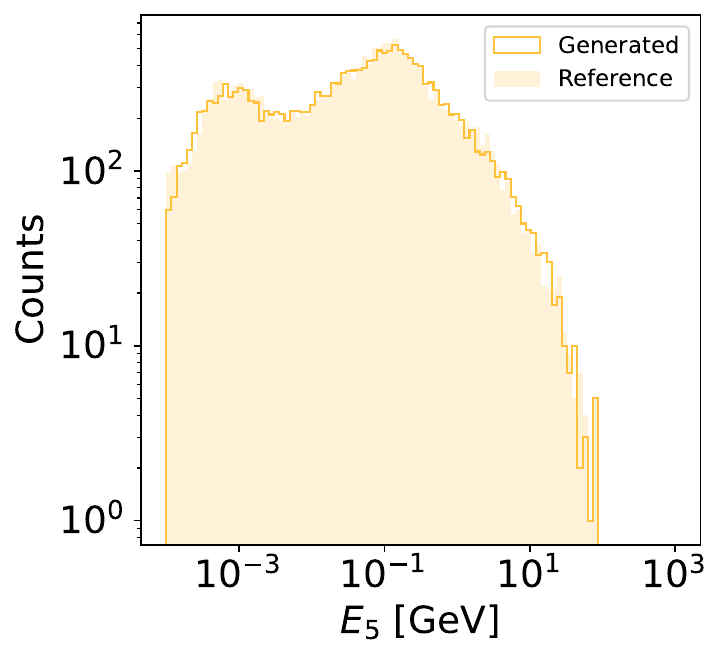}\\
\includegraphics[width=0.66\columnwidth]{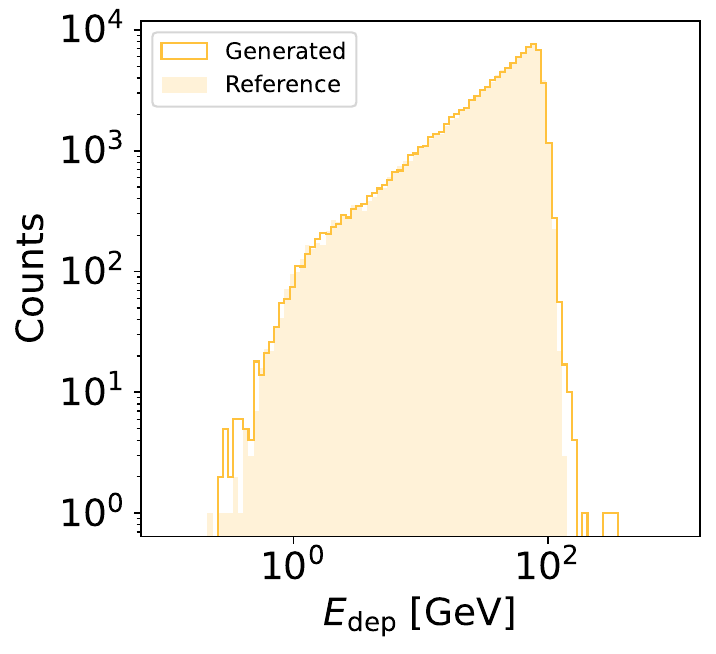}\includegraphics[width=0.66\columnwidth]{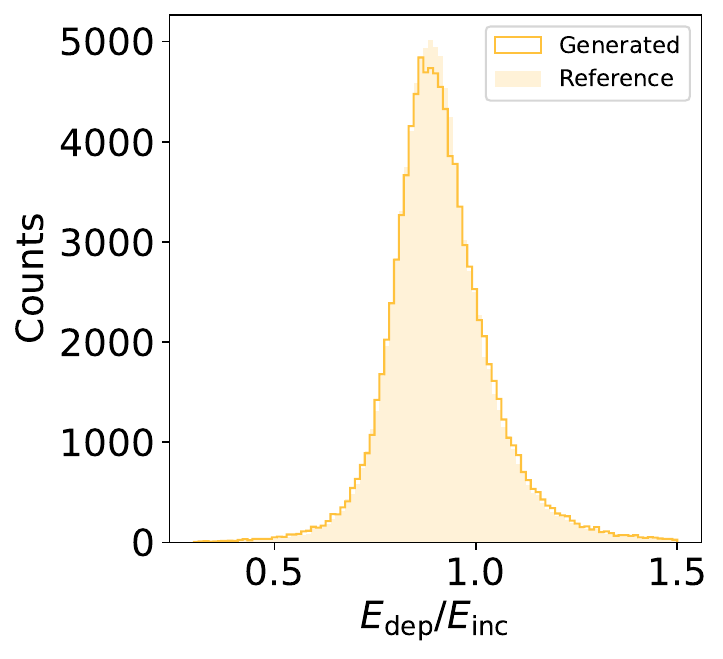}

\caption{Generated versus reference energy distributions in our new ECAL+HCAL calorimeter setup for uniformly distributed incident energies $E_{\rm inc}$. The energy deposited in layer $i$ is denoted by $E_i$ and the energy deposited in the calorimeter is denoted by $E_{\rm dep} = \sum_{i=0}^6 E_i$. The ratio of energy deposited in the calorimeter and the incident energy of the particle is denoted by $E_\text{dep}/E_\text{inc}$.}
\label{fig:energy_dist}
\end{figure*}

In this representation, the calorimeter shower geometry is made up of voxels (volumetric pixels) and the details of the calorimeter voxel dimensions are included in Table \ref{tab:calorimeter_specs}. The energy distribution in the calorimeter for uniformly distributed $\pi^+$ incident energies is shown in Fig.~\ref{fig:energy_dist}. Note that the actual energy deposited in the sampling calorimeter is usually much smaller than the incident energy. Hence, it is common practice that the energy deposited is rescaled\footnote{In this work, the rescaling is done by dividing the actual energy deposited in the ECAL or HCAL by their respective sampling fractions.} based on the sampling fraction to ensure that the dataset contains deposited energy $E_{\rm dep}$ that is close to the incident energy $E_{\rm inc}$. For this reason, it is possible for some showers to have $E_{\rm dep}/E_{\rm inc}>1$ as shown in Fig.~\ref{fig:energy_dist}.

\begin{table}[ht]
\begin{center}
%\resizebox{\columnwidth}{!}{
%\LARGE
\begin{tabular}{|c|c|c|c|c|c|}
\hline
\begin{tabular}{c} \end{tabular} & \begin{tabular}{c}Layer \\index\end{tabular} & \begin{tabular}{c}$z$ length \\ (mm) \end{tabular} &\begin{tabular}{c}$\eta$ length \\ (mm) \end{tabular} & \begin{tabular}{c}$\phi$ length \\ (mm) \end{tabular} & \begin{tabular}{c}Number \\of voxels\end{tabular}\\
\hline\hline
\multirow{3}{*}{ECAL} &0 & 90 & 5 & 160 & $3\times96$ \\ 
\cline{2-6}& 1 & 347 & 40 & 40 & $12\times12$ \\ 
\cline{2-6}& 2 & 43 & 80 & 40 & $12\times6$ \\ \hline\hline
\multirow{3}{*}{HCAL} & 3 & 375 & 20.83 & 666.67 & $3\times96$ \\ 
\cline{2-6}& 4 & 667 & 166.67 & 166.67 & $12\times12$ \\ 
\cline{2-6}& 5 & 958 & 333.33 & 166.67 & $12\times6$ \\ \hline
\end{tabular} %}
\caption{Dimensions of a calorimeter voxel. The positive $z-$axis (radial direction in full detector) is the direction of particle propagation, the $\eta$ direction is along the proton beam axis, and $\phi$ is perpendicular to $z$ and $\eta$. For the number of voxels, the first (second) number is the number of bins in the $\phi$ ($\eta$) direction (e.g., $12\times 6$
refers to 12 $\phi$ bins and 6 $\eta$ bins).}
\label{tab:calorimeter_specs}
\end{center}
\end{table}

\subsection{Calorimeter Shower simulation with Normalizing Flows}
\label{sec:caloflow}

\cf~\cite{Krause:2021ilc,Krause:2021wez} is an approach to fast calorimeter simulation based on conditional normalizing flows. In the context of fast calorimeter simulation, \cf~ is designed to generate the voxel level shower energies $\vec{\mathcal{I}}$ conditioned on the corresponding incident energies of the showers $E_\text{inc}$ denoted by $p(\vec{\mathcal{I}}|E_\text{inc})$. In particular, it uses a two-flow approach to perform this task: Flow-I is formulated to learn the probability density $p_1(\vec E|E_\text{inc})$ of calorimeter layer energies\footnote{The layer energy of a given calorimeter layer is the sum of all the voxel energies in that layer.} $\vec E$ conditioned on the incident energy $E_\text{inc}$, while Flow-II is designed to learn the probability density of the normalized voxel level shower energies conditioned on the calorimeter layer energies $p_2\left(\vec{\mathcal{I}}| \vec E, E_\text{inc}\right)$. 

For the regression task discussed in this section, we found that having layer energy information was sufficient and adding voxel level information did not significantly increase the regression performance.  This is consistent with Ref.~\cite{Acosta:2023nuw}, which found that increasing longitudinal segmentation is more effective than increasing transverse segmentation (equivalent to adding voxel level information) in improving the energy resolution. As we solely used $\vec E$ as input, only Flow-I of \cf~was required to perform the calibration task.

In our present dataset, $\vec{E}=(E_0, E_1, E_2, E_3, E_4, E_5)$ is six-dimensional, twice the dimensionality found in previous versions of the dataset. Hence, there were some minor modifications made to the original \cf~primarily to deal with different dimensionality of the new dataset. Nevertheless, the main \cf~algorithm remains the same. The details of the architecture and training are outlined in Appendix~\ref{sec:arch_training}. It is known that using an ensemble of generative models generally helps to improve generation performance at the distribution level (e.g. Ref.~\cite{butter2021amplifying}) and would ideally be done for the task of fast calorimeter surrogate modelling. Hence, we opted to train two ensembles of 10 independent Flow-I models each. In particular, 10 models are trained on uniformly distributed $E_{\rm inc}$, and the other 10 models are trained on log-uniformly distributed $E_{\rm inc}$.

When generating samples with an ensemble of models, each model in the ensemble is used to generate an equal fraction of the total number of samples. We show in Fig.~\ref{fig:energy_dist} that an ensemble of Flow-I models is able to generate energy distributions that agree reasonably well with the reference distributions with some slight mismatch in tails and at places with sharp changes in the reference distributions.

Once \cf~is trained, it can be repurposed for the calibration of particle incident energy without any additional retraining. For each shower, the predicted particle incident energy corresponds to the incident energy which maximizes the likelihood function estimated by \cf. The ``full" likelihood function of the ensemble of Flow-I models is taken to be mean of the likelihood functions from each of the models in the ensemble. For the remainder of this paper, this ``full" likelihood is what we use to perform the MLE-based calibration.

\subsection{Prior-independent and less biased calibration}
\label{sec:bias}
Using the dataset described in Sec.~\ref{sec:dataset} to train our models, we predict the incident energy of the incoming pion $E_\text{inc}$ given the energy deposited in the six calorimeter layers $\vec E$. To test the prior dependence of these methods, we also compared the performance of these methods when trained on showers with uniformly distributed $E_\text{inc}$ versus log-uniformly distributed $E_\text{inc}$ in the range [1,100] GeV. 

Our chosen direct regression model is a dense neural network (DNN) with two fully connected hidden layers, each comprising 256 nodes with ReLU activation functions. The direct regression DNN is implemented in \textsc{PyTorch} and optimized with \textsc{Adam}~\cite{kingma2014adam} using a batch size of 200 and 200 epochs. We found that using an ensemble of DNNs helped to reduce the uncertainty of the calibration. Like for \cf~, we opted to have two ensembles, each with 10 independent DNNs. In particular, 10 models are trained on uniformly distributed $E_{\rm inc}$, and the other 10 models are trained on log-uniformly distributed $E_{\rm inc}$. For each shower, the predicted incident energy of the ensemble is then taken as the mean predicted incident energy of all the individual DNNs in the ensemble. Each DNN has $\sim$134k model parameters, while each NF has $\sim$107k model parameters.

\begin{figure*}[!ht]
    \includegraphics[width=0.65\columnwidth]{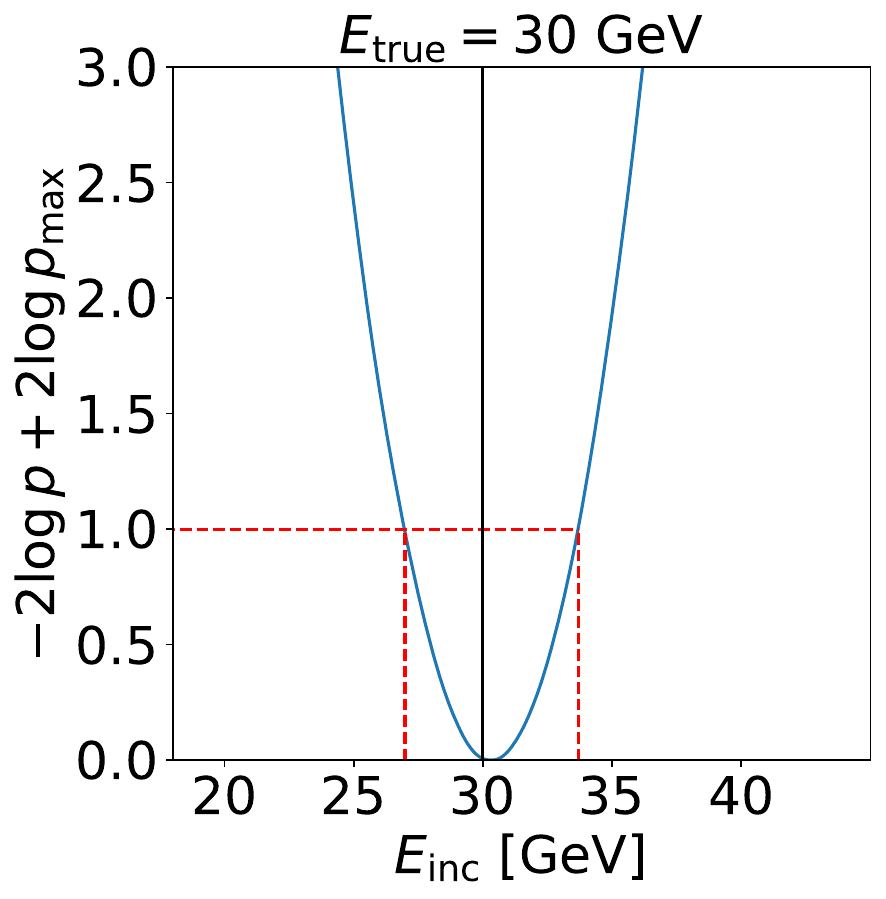}\includegraphics[width=0.65\columnwidth]{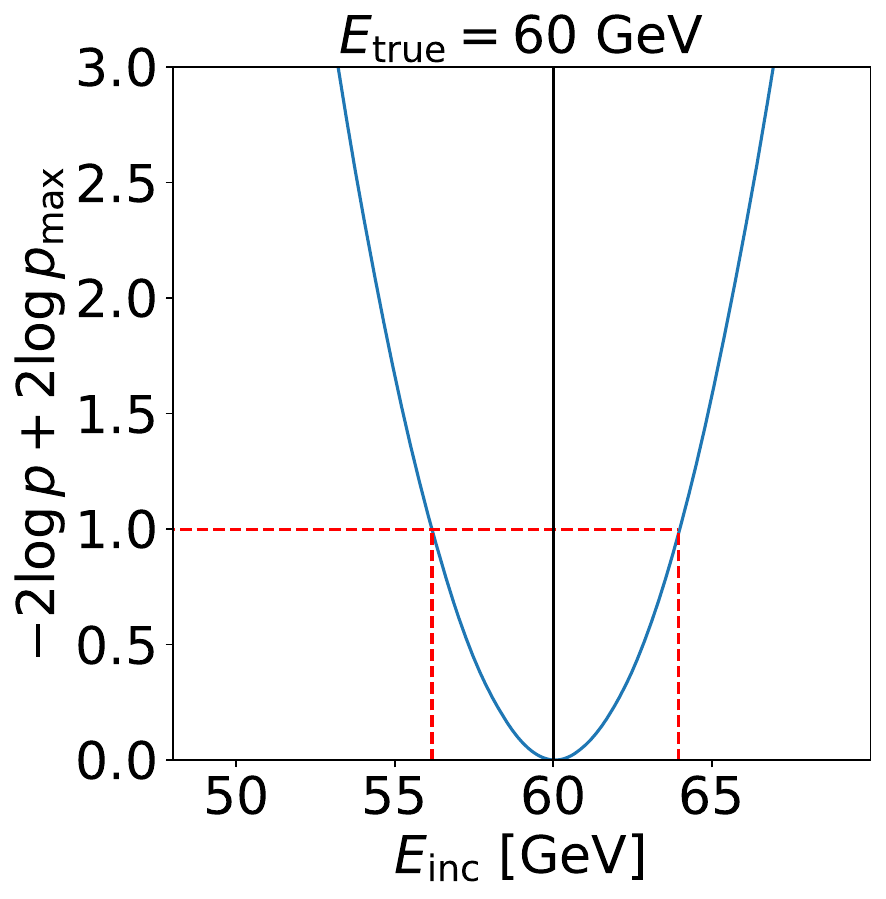}\includegraphics[width=0.65\columnwidth]{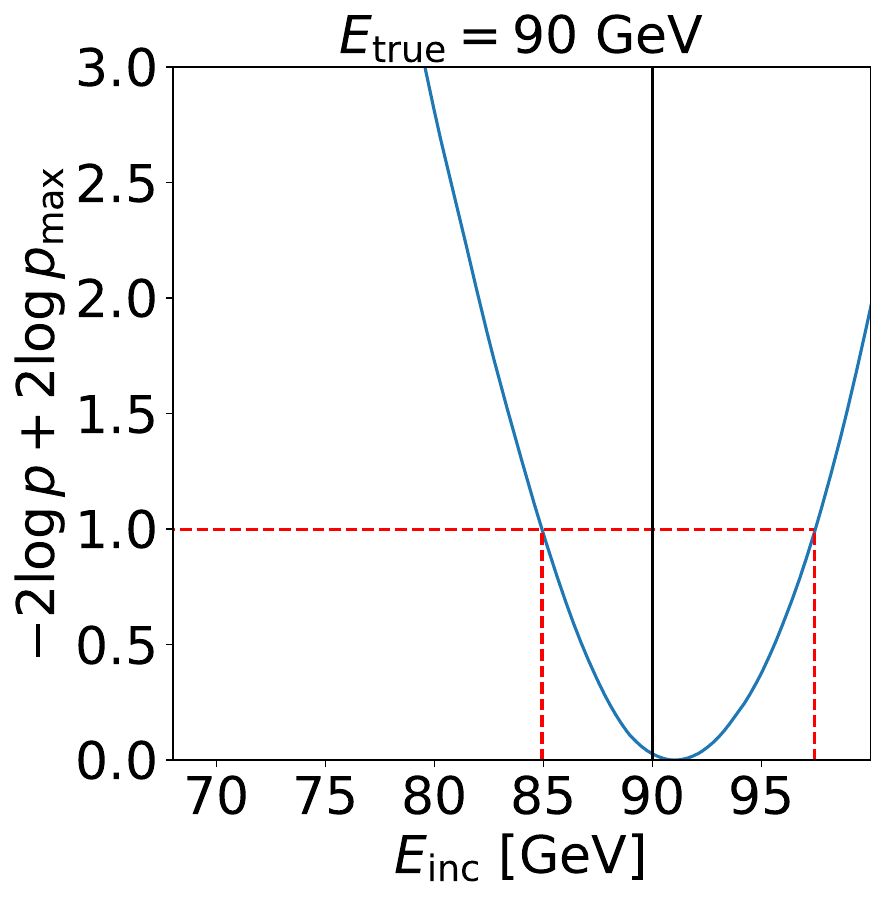}
    \caption{Plot of likelihood evaluated using \cf~for a given calorimeter shower originating from a 30 GeV pion (left), 60 GeV pion (middle) and 90 GeV pion (right), respectively. In each plot, the true incident energy $E_{\rm true}$ is shown by the vertical black line, and the evaluated likelihood by the NF is shown in blue. The lower (upper) bounds of the 68\% confidence interval about the MLE are shown by the red vertical dashed lines.}
    \label{fig:MLE_eg}
\end{figure*}

In Fig.~\ref{fig:MLE_eg}, we show examples of the likelihood evaluated by \cf~for three given calorimeter showers that originate from a 30 GeV pion, 60 GeV pion and 90 GeV pion, respectively. Specifically, the difference between $-2\log p(\vec E|E_{\rm inc})$ and $-2\log p_{\rm max} \equiv -2\log p(\vec E|E_{\rm inc}) \vert_{E_{\rm inc} = E_{\rm pred}}$ is plotted for a range of $E_\text{inc}$, where $E_{\rm pred}$ is the MLE prediction from \cf. Note that the value of $E_{\rm pred}$ corresponds to the minimum of the blue curve in each of the three plots in Fig.~\ref{fig:MLE_eg}.

To evaluate the bias, the mode of the $E_{\rm pred}/E_{\rm true}$ distribution has to be computed for each fixed $E_{\rm true}$. In this study, the mode is estimated using kernel density estimation (KDE)\footnote{We used the \texttt{scipy.stats.gaussian\_kde}~\cite{2020SciPy-NMeth} function to perform the KDE.} and the uncertainty of the estimate is determined by bootstrapping~\cite{10.1214/aos/1176344552}. To obtain the mode of the $E_{\rm pred}/E_{\rm true}$ distribution at each fixed $E_{\rm true}$, the steps we took are as follows:

\begin{enumerate}
    \item \textbf{Obtain predicted energies:} \\ For a given $E_{\rm true}$ (e.g., 30 GeV), collect $N$ values of $E_{\rm pred}$ predicted by the model. Here $N$ is the number of showers in the evaluation dataset for a given fixed $E_{\rm true}$. For each $E_{\rm pred}$ value, form the ratio $E_{\rm pred}/E_{\rm true}$. 
    \item \textbf{Bootstrap resampling:} \\Draw with replacement $N$ samples from $N$ values of $E_{\rm pred}$.
    \item \textbf{Kernel density estimation (KDE):}\\ Perform kernel density estimation of the drawn samples with kernel bandwith determined using Scott's rule~\cite{scott}.
    \item \textbf{Identify the mode:}\\ Identify the position of the mode of the estimated density. This value represents the mode of the distribution for that particular bootstrap sample.
    \item \textbf{Repeat and accumulate:}\\ Repeat steps 2-4 for a total of 20 times. Store the mode estimate from each iteration.
    \item \textbf{Compute final estimate and uncertainty:} \begin{itemize}
        \item \textit{Final mode estimate:} Take the mean of the 20 mode values 
        obtained in the previous step.
        \item \textit{Uncertainty:} Use the standard deviation of those 20 mode 
        estimates as the uncertainty on the final mode.
    \end{itemize}
\end{enumerate}

\begin{figure}[!ht]
    \centering
    \includegraphics[width=\columnwidth]{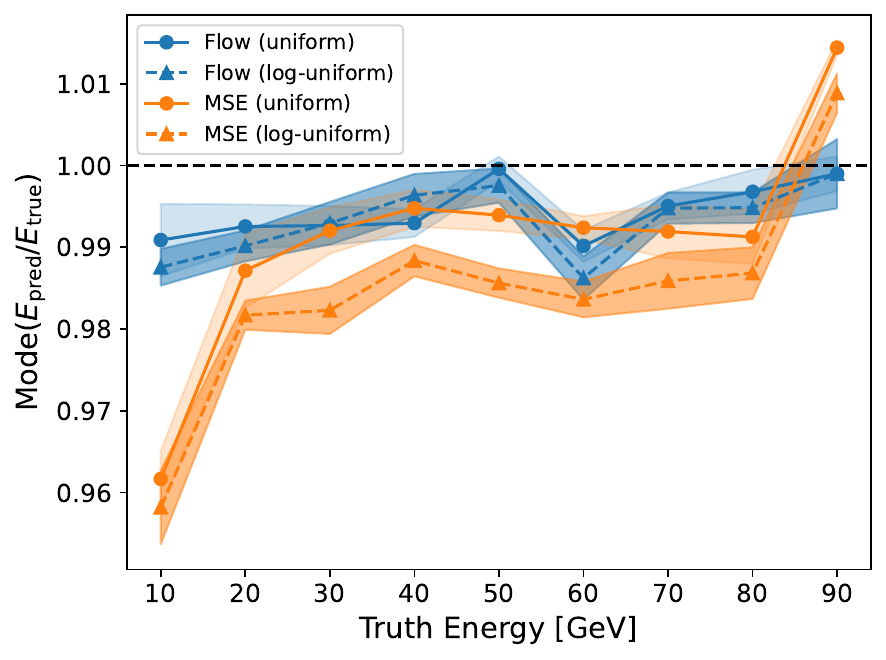}
    \caption{Plot of the mode of $E_{\rm pred}/E_{\rm true}$ distribution, denoted by ``Mode$(E_{\rm pred}/E_{\rm true})$", and the corresponding uncertainty at different fixed values of $E_{\rm true} \in \{10,20,30,40,50,60,70,80,90\}$ GeV. Results for models trained on (log-)uniformly distributed $E_\text{inc}$ are shown in solid (dashed) lines.}
    \label{fig:bias_prior_dep}
\end{figure}

  The mode estimation is performed for both the uniform and log-uniform cases. In Fig.~\ref{fig:bias_prior_dep}, we show the the mode of $E_{\rm pred}/E_{\rm true}$ distribution at $E_{\rm true} \in \{10,20,30,40,50,60,70,80,90\}$ GeV for the different regression methods. Here we used nine evaluation datasets, each with 100k $\pi^+$ showers at one of the nine fixed $E_{\rm true}$ values.

 We observe that the flow-based calibration is generally less biased (closer to unity) compared to the MSE-based calibration. The larger bias of the MSE-based calibration is clearly seen at the edges of the $E_{\rm true}$ range. We note that it may be possible to address this limitation with e.g.~domain continuation~\cite{CMS:2014afl}. Domain continuation has been shown to improve regression near the high energy edge of the region-of-interest (ROI) by extending the training data to include even larger energies beyond the ROI. Close to the small energy edge of the ROI, one may need to extrapolate the training data into the unphysical, negative energy region. The regressed values lying outside the ROI are discarded/ignored in the analysis and this results in lower efficiency. In contrast, our method does not result in regressed values being discarded.

Furthermore, we see that the MSE-based calibration corresponding to the log-uniform prior is generally more biased than the calibration corresponding to the uniform prior at intermediate energies.
 
 The prior independence of the flow-based calibration is confirmed by observing consistent biases of \cf~trained on uniform and log-uniform priors. For the MSE-based calibration, there is prior dependence at many of the $E_{\rm true}$ values used in the evaluation as seen from the difference in the biases corresponding to the uniform and log-uniform priors. For \cf~, the origin of the dip at $E_{\rm true}=60$ GeV is unclear. However, we observe that the biases of \cf~trained on uniform and log-uniform priors agree with each other.

\subsection{Resolution estimation}
\label{sec:resolution}
MLE-based calibration methods, such as the NF and Gaussian Ansatz, are able to provide estimates of the per-shower resolution and this allows these methods to go beyond a point estimate by providing per-event resolutions (i.e. per-shower resolutions in calorimeter context). This gives MLE-based calibration an important advantage over direct regression. Unlike the Gaussian Ansatz, the NF serves as a generative model, presenting a notable advantage over the Gaussian Ansatz: a NF utilized for surrogate modeling can subsequently be repurposed for calibration tasks without additional retraining.

\begin{figure}[!ht]
    \centering
    \includegraphics[width=\columnwidth]{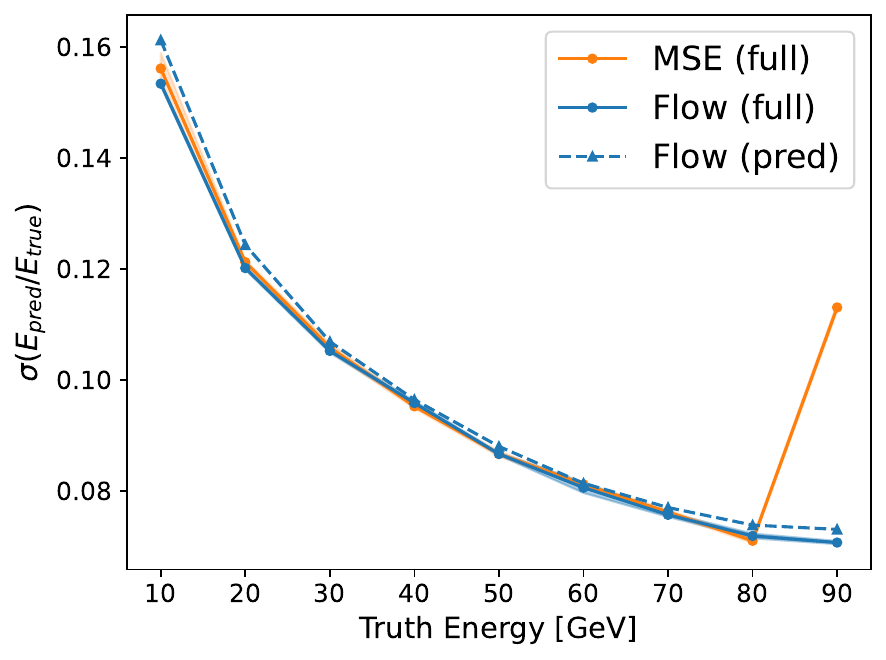}
    \caption{Plot of full resolution (solid) and mean predicted per-shower resolution (dashed) at $E_{\rm true} \in \{10,20,30,40,50,60,70,80,90\}$ GeV.}
    \label{fig:resolution}
\end{figure}

Using \cf, we perform per-shower resolution estimation by computing the two $E_{\rm inc}$ values corresponding to the lower (upper) bound of 68\% confidence interval of the MLE prediction. Equivalently, these are the two $E_{\rm inc}$ values for which $-2\log p + 2 \log p_{\rm max} =1$ as shown by the vertical dashed lines in Fig.~\ref{fig:MLE_eg}. The per-shower resolution prediction is then the average difference of the two $E_{\rm inc}$ values from $E_{\rm pred}$. The full resolution is computed as half of the 68\% confidence interval about the mode of the $E_{\rm pred}/E_{\rm true}$ distribution. In Fig.~\ref{fig:resolution}, we compare the mean predicted per-shower resolution and the full resolution at each of the nine fixed $E_{\rm true}$ from the NF. The results in this section were produced using the NF trained on showers with uniformly distributed incident energies. We observe that the predicted resolution from the NF closely matches the full resolution. Also, we found that the uncertainty in the determination of the full resolution to be small and almost unnoticeable in Fig.~\ref{fig:resolution}.

Furthermore, we found that for most values of $E_{\rm true}$, the MSE-based calibration obtains a full resolution that closely matches that from the NF. However, it is unable to provide per-shower resolutions. The large full resolution of the MSE-based calibration at $E_{\rm true}=90$ GeV is due to edge effects of the $E_{\rm inc}$ training data distribution.

\begin{figure*}[!ht]
    \includegraphics[width=0.7\columnwidth]{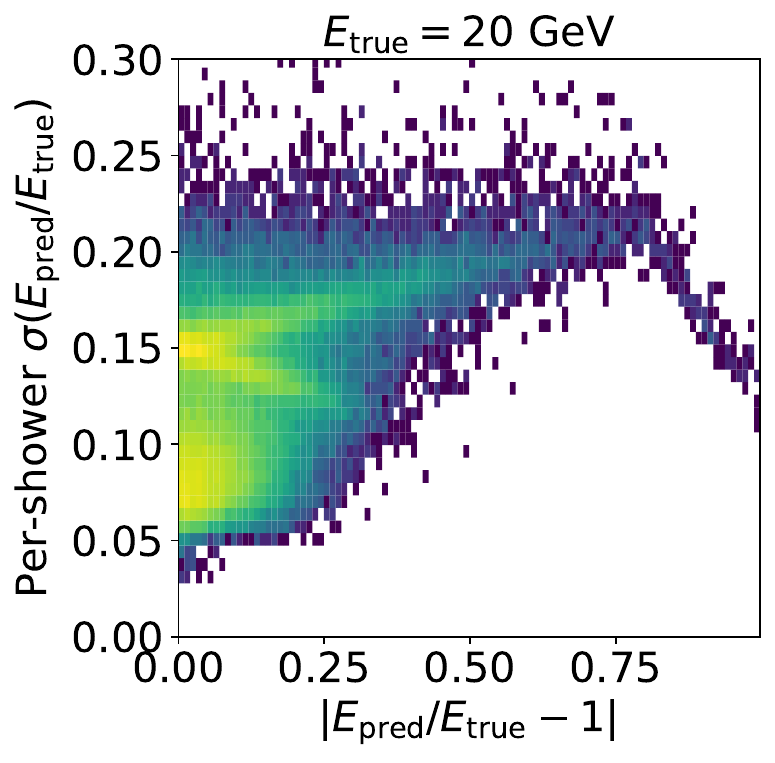}\includegraphics[width=0.7\columnwidth]{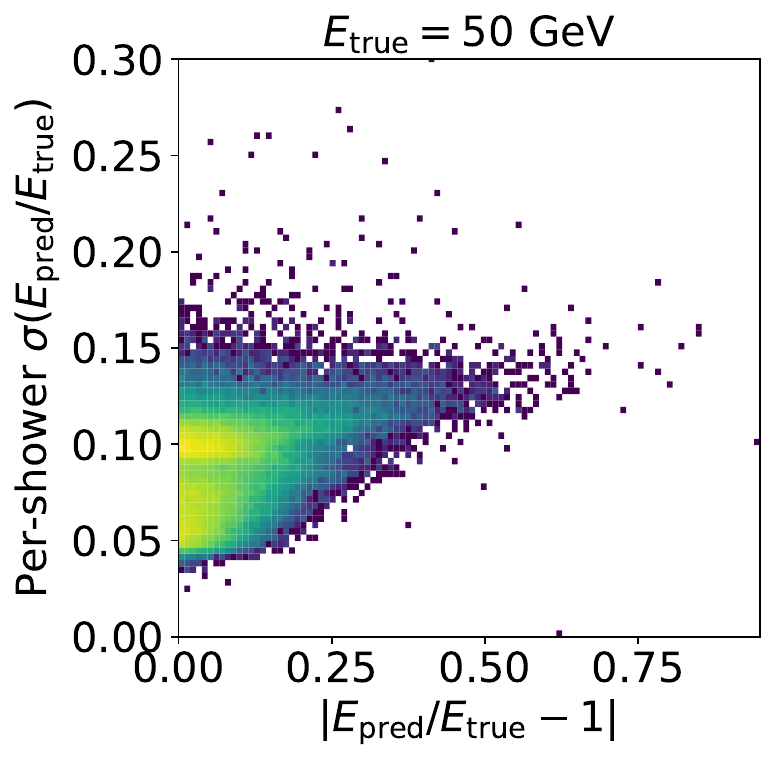}\includegraphics[width=0.7\columnwidth]{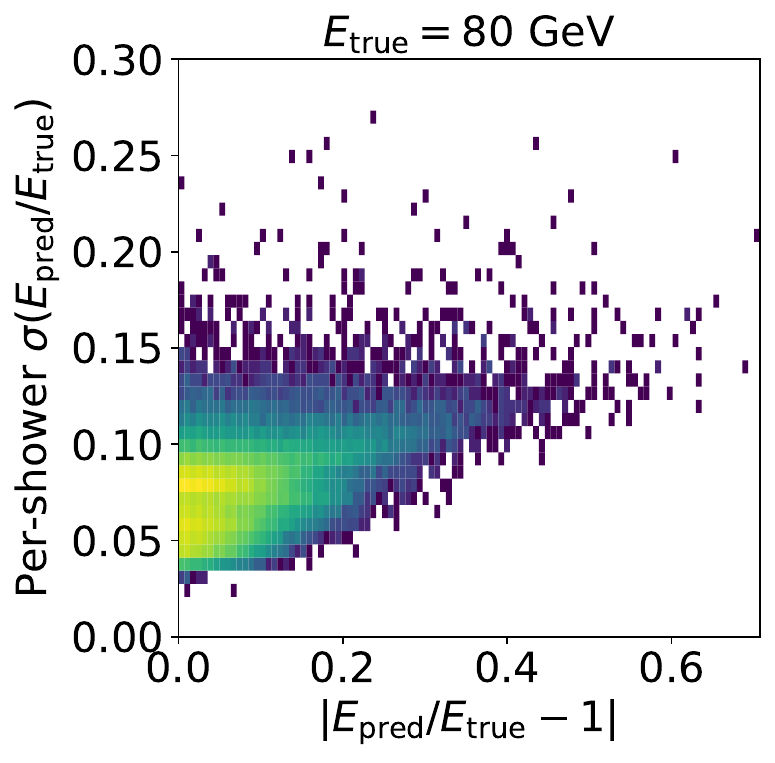}
    \caption{2D histograms of $\abs{E_{\rm pred}/E_{\rm true}-1}$ versus per-shower resolution at $E_{\rm true}=20$ GeV (left), $E_{\rm true}=50$ GeV (middle) and $E_{\rm true}=80$ GeV (right).}
    \label{fig:per_shower}
\end{figure*}

Next we show that the per-shower resolution is a reliable guide to the calibration accuracy. In Fig.~\ref{fig:per_shower}, we plot 2D histograms of $\abs{E_{\rm pred}/E_{\rm true}-1}$ versus the per-shower resolution at three different $E_{\rm true}$ values. We observe that showers with small per-shower resolution tend to have small $\abs{E_{\rm pred}/E_{\rm true}-1}$, while those with large per-shower resolutions appear to be evenly distributed across different values of $\abs{E_{\rm pred}/E_{\rm true}-1}$. In other words, a smaller per-shower resolution is indicative of a more accurately calibrated shower. For $E_{\rm true} = 20$ GeV, there is a drop in the per-shower resolution for $\abs{E_{\rm pred}/E_{\rm true}-1}\gtrsim 0.75$ where the NF is relatively confident that the shower originates from an $E_{\rm inc}>E_{\rm true}$. Nevertheless, we note that the per-shower resolution in this range of $\abs{E_{\rm pred}/E_{\rm true}-1}$ is still larger than that at low values of $\abs{E_{\rm pred}/E_{\rm true}-1}$.

Since the NF is able to access the complete likelihood, this enables it to capture non-Gaussian\footnote{In principle, the Gaussian Ansatz~\cite{Gambhir:2022gua,Gambhir:2022dut} should also be able to capture non-Gaussian components of the resolution. However, this has yet to be shown in a worked example. Previous works~\cite{Gambhir:2022dut,Gambhir:2024tgs} only attempted to compute the Gaussian resolution. Also, the Gaussian Ansatz is trained in a different way (see Ref.~\cite{Gambhir:2022gua}) compared to the NF and is more sensitive to hyperparameter choices.} components of the resolution. We quantify this by defining an asymmetry observable based on the upper and lower bounds of the 68\% confidence interval of the MLE prediction. In particular, the asymmetry is computed as the ratio $\frac{\Delta_L}{\Delta_U}$, where $\Delta_L$ ($\Delta_U$) is the difference between the lower (upper) bound of the 68\% confidence interval and the MLE. 

\begin{figure}
    \centering
    \includegraphics[width=0.9\columnwidth]{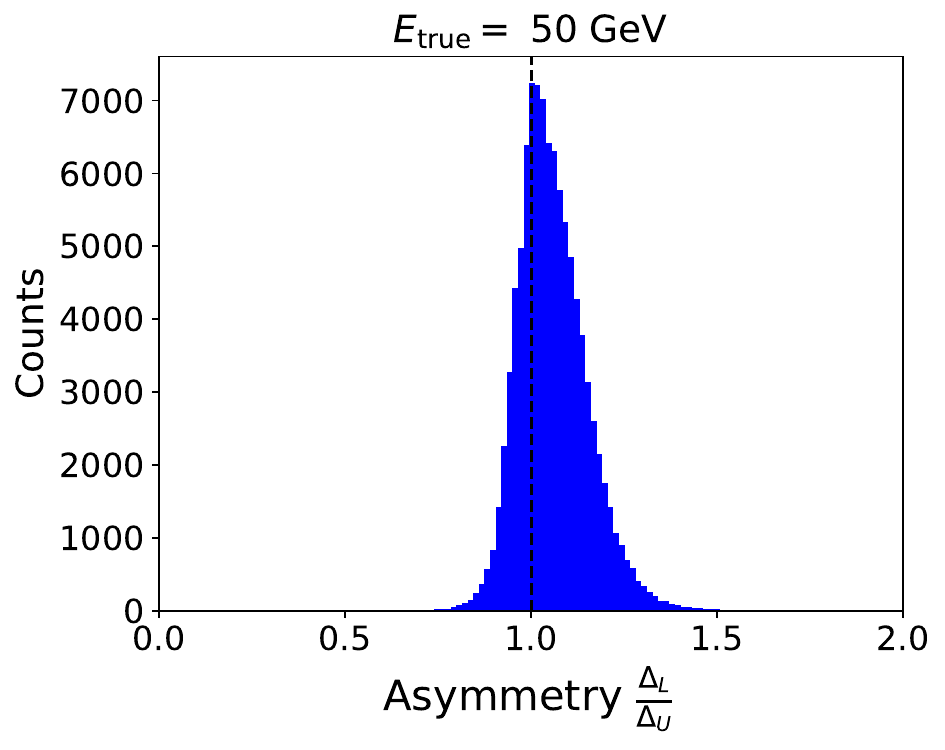}
    \caption{Histogram of asymmetry $\frac{\Delta_L}{\Delta_U}$ predicted by \cf~for 100k showers with $E_{\rm true} = 50$ GeV.}
    \label{fig:asymmetry}
\end{figure}

\begin{figure}
    \centering
    \includegraphics[width=0.9\columnwidth]{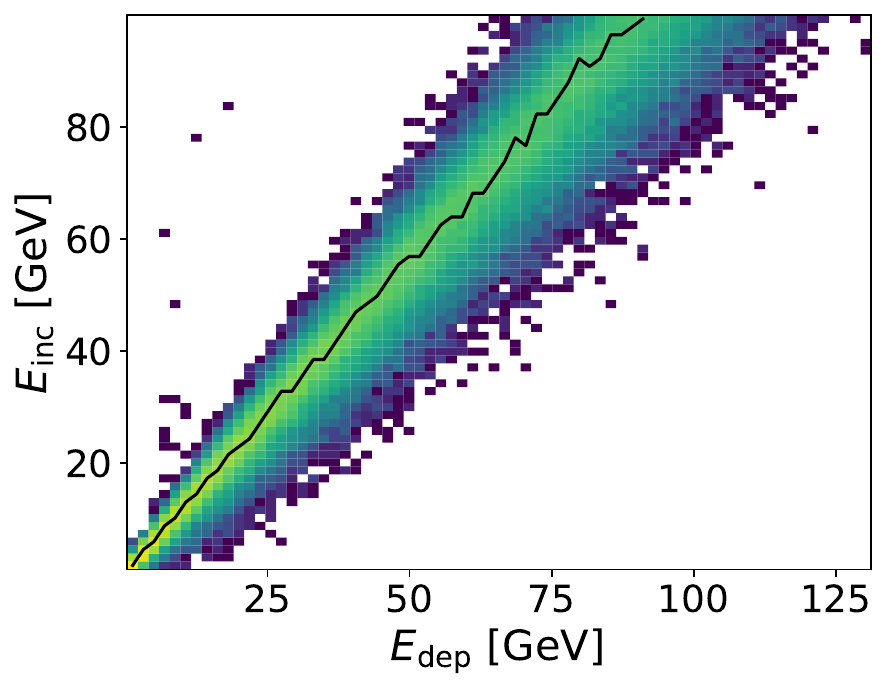}
    \caption{2D histogram of the deposited energy $E_{\rm dep}$ versus $E_{\rm inc}$ for 100k showers with uniformly distributed $E_{\rm inc}$. A solid line is drawn across the mode of $E_{\rm inc}$ at each $E_{\rm dep}$ to guide the eye.}
    \label{fig:E_asymm}
\end{figure}

An example is included in Fig.~\ref{fig:asymmetry} which shows the per-shower resolution asymmetry predicted by the NF for 100k showers with $E_{\rm true} = 50$ GeV. We observe that $\frac{\Delta_L}{\Delta_U}$ is largely greater than unity. This implies that for most showers, lower $E_{\rm inc}$ values tend to have larger likelihoods. 
While one might suspect that this resolution asymmetry is merely an artifact of the flow, we observe in Fig.~\ref{fig:E_asymm} that the distribution of $E_{\rm inc}$ at fixed vertical slices of $E_{\rm dep}$ is also asymmetric towards lower values of $E_{\rm inc}$. The particular shape of the 2D histogram in Fig.~\ref{fig:E_asymm} is due to the nature of the $\pi^+$ showers. At fixed $E_{\rm inc}$, fully electromagnetic events will have larger $E_{\rm dep}$, since all the energy is deposited via electromagnetic interactions. On the other hand, fully hadronic events will have  significantly lower $E_{\rm dep}$ due to energy carried away invisibly by particles like neutrons and neutrinos produced in nuclear interactions. The $\pi^+$ showers considered in this work contain both electromagnetic and hadronic components, so on a shower-by-shower basis, the $E_{\rm dep}$ can range anywhere between these two extremes of a fully electromagnetic shower and a fully hadronic shower~\cite{RevModPhys.75.1243}. This results in a tail at high energies in the $E_{\rm dep}$ distribution for fixed $E_{\rm inc}$, presumably because the typical $\pi^+$ shower is mostly hadronic in nature. Thinking of $p(E_{\rm dep}|E_{\rm inc})$ as a proxy for the likelihood suggests that the resolution asymmetry is not merely an artifact of the flow but a genuine characteristic to be expected.

\section{Conclusions}
Using \cf~as an example, we show that normalizing flow-based deep generative models can be repurposed for prior-independent detector calibration.

We compared the calibration performance of \cf~and a DNN direct regression model on $\pi^+$ calorimeter showers from our new sampling calorimeter dataset~\cite{du_2024_11073232}, and found that the direct regression method is clearly more biased relative to the flow-based calibration method. This highlights the advantage of utilizing prior-independent MLE-based calibration for tasks such as particle energy regression.

We demonstrate a second advantage of using MLE-based calibration methods over direct regression by estimating per-shower resolutions. The average estimated per-shower resolution obtained from \cf~closely aligns with the full resolution. In contrast, the direct regression approach yields only point estimates and lacks the capacity to make per-shower resolution predictions. For \cf, we found that smaller per-shower resolutions tend to coincide with more accurately calibrated showers. This is evidence that the per-shower resolutions from \cf~are reliable and have the potential to be leveraged for improved calibration. Notably, utilizing a NF for calibration grants access to the complete resolution function. This characteristic enables the NF to capture asymmetries in the resolution. In future work, it would be interesting to study how the per-shower resolution information obtained by the NF can be utilized to further improve the calibration.
\label{sec:conclusions}

\section*{Data and code availability}
The datasets used in this study can be found at Ref.~\cite{du_2024_11073232} and the software to generate these datasets are located at \href{https://github.com/hep-lbdl/CaloGAN/tree/two\_layer}{https://github.com/hep-lbdl/CaloGAN/tree/two\_layer}. The machine learning software is at \href{https://github.com/Ian-Pang/regression\_with\_CF}{https://github.com/Ian-Pang/regression\_with\_CF}. 

\section*{Acknowledgements}
We would like to thank Rikab Gambhir and Jesse Thaler for the helpful discussions related to the Gaussian Ansatz model and for feedback on the manuscript. CK would like to thank the Baden-W\"urttemberg-Stiftung for financing through the program \textsl{Internationale Spitzenforschung}, pro\-ject \textsl{Uncertainties – Teaching AI its Limits} (BWST\_IF2020-010). IP and DS are supported by the U.S. Department of Energy (DOE), Office of Science grant DOE-SC0010008 and HD, VM, and BN are supported by the DOE under contract DE-AC02-05CH11231.

\appendix
\section{Architecture and training}
\label{sec:arch_training}
Here we briefly describe the architecture and training procedure used for \cf~(see Refs.~\cite{Krause:2021ilc,Krause:2021wez} for more details). There are some differences compared to the implementation in the original \cf~papers~\cite{Krause:2021ilc,Krause:2021wez}, but most of the main algorithm remains the same. One main difference is that only Flow-I is used in this study.

The flows used in this work are Masked Autoregressive Flows (MAFs)~\cite{papamakarios2017masked} with compositions of Rational Quadratic Splines (RQS)~\cite{durkan2019neural} as transformations. The RQS transformations are parametrized using neural networks known as MADE blocks~\cite{germain2015made}. Identical flow architectures are used in each of the two cases with uniformly and log-uniformly distributed $E_{\rm inc}$. Each flow consists of six MADE blocks, each with two hidden layers of 64 nodes. The RQS transformations are defined with 8 bins and a tail bound of 14.

The incident energy of the incoming photon is preprocessed as 

\begin{equation}
E_\text{inc} \to \log_{10} (E_\text{inc}/10 \ \text{GeV})\,.
\end{equation}

The layer energies are preprocessed as 
\begin{equation}
    E_i \to 2\left(\log_{10}(E_i+1 \ \text{keV})-1\right)\,.
\end{equation}

The index $i$ denotes the layer number. In the original \cf, a different preprocessing was used for the layer energies $E_i$ in Flow-I where $E_i$ were transformed to unit-space (see \cite{Krause:2021ilc}). 
 
Uniform noise in the range [0,0.1] keV was applied to the voxel energies during training and evaluation. The same range was used for the pion dataset in~\cite{Krause:2022jna}. The range of uniform noise used in the original \cf~\cite{Krause:2021ilc,Krause:2021wez} was [0,1] keV. The addition of noise was found to prevent the flow from fitting unimportant features. The training of flows in this work is optimized using independent \textsc{Adam} optimizers~\cite{kingma2014adam}. The flows were trained by minimizing $-\log p_1(\vec E|E_\text{inc})$ for 150 epochs with a batch size of 200. The initial learning of $10^{-4}$ was chosen and a multi-step learning schedule was used when training the flow which halves the learning rate after each selected epoch milestone during the training.

\bibliographystyle{JHEP}
\bibliography{HEPML,other-refs}

\end{document}